\documentclass[showpacs,amsmath,amssymb,aps,10pt,reprint,superscriptaddress]{revtex4-1}
\usepackage{graphicx}
\usepackage{bm}
\usepackage[breaklinks=true,colorlinks=true,linkcolor=blue,urlcolor=blue,citecolor=blue]{hyperref}
\usepackage{graphics}
\usepackage{dcolumn}
\usepackage{amsmath,amssymb}
\usepackage{mathdots}
\usepackage{natbib}
\usepackage{soul,color}
\usepackage{epstopdf}

\begin{document}

\title{Pinning effects on self-heating and flux-flow instability\\ in superconducting films near $T_{c}$}

\author{Valerij~A.~Shklovskij}
\affiliation{Physics Department, V. Karazin Kharkiv National University, 61077 Kharkiv, Ukraine}
\author{Anastasiia P. Nazipova}
\affiliation{Faculty of Science, P. J. \v Saf\'arik University in Ko\v sice, SK-04001 Ko\v sice, Slovakia}
\affiliation{Physics Department, V. Karazin Kharkiv National University, 61077 Kharkiv, Ukraine}
\author{Oleksandr~V.~Dobrovolskiy}
\affiliation{Physikalisches Institut, Goethe University, 60438 Frankfurt am Main, Germany}
\affiliation{Physics Department, V. Karazin Kharkiv National University, 61077 Kharkiv, Ukraine}

\date{\today}

\begin{abstract}
The effect of pinning on self-heating triggering the Larkin-Ovchinnikov (LO) flux-flow instability (FFI) in superconducting thin films is theoretically investigated. The problem is considered relying upon the Bezuglyj-Shklovskij (BS) generalization of the LO theory, accounting for a finite heat removal from the quasiparticles at temperature $T^\ast$ to the bath at temperature $T_0$. The FFI critical parameters, namely the current density $j^{\ast}$, the electric field $E^{\ast}$, the power density $P^{\ast}$, and the vortex velocity $v^{\ast}$ are calculated and graphically analyzed as functions of the magnetic field and the pinning strength. With increasing pinning strength at a fixed magnetic field value $E^{\ast}$ \emph{decreases}, $j^{\ast}$ \emph{increases}, while $P^{\ast}$ and $T^{\ast}$ \emph{remain practically constant}. Vortex pinning may hence be the cause for eventual discrepancies between experiments on superconductors with \emph{strong pinning} and the generalized LO and BS results.
\end{abstract}

\pacs{74.25.F-, 74.25.Wx, 74.25.Qt, 74.40.De}

\maketitle

\section{Introduction }

It is well known that in superconducting films a rather strong dc transport current in a perpendicular magnetic field $B$ causes a motion of Abrikosov vortices thus leading to a nonzero, $B$-dependent resistivity. If vortex pinning is negligibly weak, the flux-flow resistivity is measurable even at small transport currents. In the flux-flow regime, the current-voltage curve (CVC) of a film is linear with
\begin{equation}
    \label{1}
    j=\sigma_{f}E,
\end{equation}
where $j$ is the current density, $E$ is the longitudinal electric field strengh, $\sigma_{f}=\sigma_{f}(B,T)=\sigma_{n}H_{c2}(T)/B$ is the temperature-dependent Bardeen-Stephen \cite{Bar65prv} flux-flow conductivity. Here $\sigma_{n}$ is the normal metal film conductivity and $H_{c2}(T)$ is the upper critical field.

However, Larkin and Ovchinnikov (LO) showed theoretically \cite{Lar75etp,Lar86inb} that the flux-flow regime at $T$ close to the superconducting transition temperature $T_{c}$ becomes unstable at large current densities $j\simeq j^\ast$ which, still, are \emph{by far smaller} than the \emph{depairing} current density. It is this instability current density $j^\ast$ which sets the real upper limit for the current a superconductor can carry without dissipation.

In general, various mechanisms were suggested to explain voltage jumps in the CVCs of superconductors. To name a few, these are a thermal runaway effect due to Joule heating \cite{Xia99prb,Gon03prb}, formation of localized normal domains which appear in places of maximum current due to an inhomogeneous current distribution \cite{Vod07prb,Sil10prl}, crystallization of the vortex system \cite{Kos94prl,Kok04prb}, phase-slip centers or lines \cite{Siv03prl,Ber09prb}, and the Kunchur hot-electron instability \cite{Kun02prl,Bab04prb} observed at $T \lesssim 0.5T_c$ and related to thermal effects diminishing the superconducting order parameter, thus leading to an expansion of the vortex cores.

In the present work we will deal with the LO flux-flow instability at $T\simeq T_c$ relying upon the well-accepted LO instability scenario: The electric field arising due to the vortex motion accelerates quasiparticles within the vortex cores. This process continues as long as the quasiparticles energy becomes sufficient for their escape. If the time of the quasiparticle energy relaxation $\tau_{\varepsilon}$ and the respective diffusion length in ``dirty'' films $l_{\varepsilon}= \sqrt{D\tau_{\varepsilon}}$ substantially exceeds the core size of the order of the coherence length $\xi$, then the excitations can leave the core. Here $D=lv_{F}/3$, where $l$ is the electron mean free path and $v_{F}$ is the Fermi velocity. The escape of the quasiparticles from the core under the influence of the electric field $E$ causes the vortex core to shrink according to the LO relation
\begin{equation}
    \label{2}
    \xi^{2}(v)=\xi^{2}(0)/[1+(v/v^{\ast})^2],
\end{equation}
where $v$ is the vortex velocity, $\xi(0)=\xi(v=0)$, and $v^{\ast}$ is the critical vortex velocity. The decrease of $\xi$ leads to a reduction of the vortex viscosity $\eta$ given by
\begin{equation}
    \label{3}
    \eta(v)=\eta(0)/[1+(v/v^{\ast})^2],
\end{equation}
and this is why the viscous force $F_{v}=\eta(v)v$ has a maximum at $v=v^{\ast}$. A further increase of $v > v^{\ast}$ causes a reduction of $F_{v}$. In turn, this leads to an even further increase of the vortex velocity and in this way results in the instability of the vortex motion.

Experimentally, for current-driven measurements at not too large magnetic fields ($B \lesssim 0.4H_{c2}$) the nonlinear resistive part of the CVC usually exhibits a jump-like voltage rise (see, e.g. \cite{Mus80etp,Kle85ltp}). According to LO, these jumps emerge from the instability in the homogeneous flux-flow at the $B$-\emph{independent} critical velocity
\begin{equation}
    \label{4}
    v^{\ast}=1.143(D/\tau_{\varepsilon})^{1/2}(1-T/T_{c})^{1/4},
\end{equation}
when the Lorentz force equals to the maximal viscous damping force $F_v$.

The values of $\tau_{\varepsilon}$ for In, Sn, and Al deduced from early experiments \cite{Mus80etp,Kle85ltp} by Eq. (\ref{4}) agreed in the order of magnitude with the theoretical estimates. However, the authors of Ref. \cite{Kle85ltp} revealed an anomalous dependence of $\tau_{\varepsilon}$ on the applied magnetic field value. Another essential discrepancy between the LO theory and experiment \cite{Kle85ltp} lies in a noticeable decrease of the instability current density $j^{\ast}$ with increasing magnetic field, whereas in the LO theory $j^{\ast}$ does not depend on $B$ at small fields ($B\ll H_{c2}$).

Later on, Bezuglyj and Shklovskij (BS) suggested \cite{Bez92pcs} that the abovementioned discrepancies between the LO theory and experiments may have a common cause, namely the quasiparticles overheating not only inside the vortex cores, but also outside them. The latter ensues in experiments due to a finite rate of heat removal from the sample. Whereas LO supposed the temperature of phonons in the sample to be independent of the electric field value, BS argued that the  phonon overheating is unavoidable since the rate of heat removal from the film always remains finite. In the BS generalization of the LO approach BS solved the linear heat balance equation
\begin{equation}
    \label{5}
    h(T_{0})(T^{\ast}-T_{0})=d\sigma(E^{\ast})(E^{\ast})^2
\end{equation}
in conjunction with the CVC extremum condition
\begin{equation}
    \label{6}
    \frac{d}{dE}[\sigma_{f}(E)E]\Big|_{E=E^{\ast}}=0.
\end{equation}
Here $h(T_{0})$ is the heat transfer coefficient from the quasiparticles at the critical temperature $T^{\ast}$ to the bath with the temperature $T_0$, $E^*$ is the critical electric field, and $d$ is the film thickness. Two regimes with respect to magnetic field values have been revealed, separated by the critical (overheating) field $B_T$ introduced by BS. Namely, when $B\ll B_T$ quasiparticles overheating is negligible and $v^{\ast}$ is given by the LO formula (\ref{4}). For $B\gg B_T$ overheating becomes essential and the measured value of $v^{\ast}$ becomes $B$-dependent: At small fields it noticeably decreases with increasing field. An excellent agreement with the BS approach has been confirmed experimentally for both, low-temperature \cite{Vol92fnt,Per05prb} and high-$T_c$ \cite{Xia99prb,Xia98prb} superconducting thin films.

Nevertheless, both the LO and BS approaches to explain the nonlinear CVC behavior as caused by the flux-flow instability (FFI) capture \emph{no pinning} in the physical picture. In reality, however, vortex pinning is omnipresent in superconducting films, and recent experiments on nanopatterned superconductors aimed at revealing its effect on the FFI critical parameters \cite{Leo10pcs,Sil12njp,Gri12apl,Gri15prb,Leo16prb,Dob17arx}. While a theoretical account for FFI at lower temperatures ($T \lesssim T_c/2$) has recently been given \cite{Shk17arx} and already allowed us to fit experimental data to the derived analytical expressions \cite{Dob17arx}, the respective generalization of the BS approach at temperatures $T\backsimeq T_c$ has not been available so far.

To understand qualitatively the pinning effect on the FFI critical parameters, two limiting cases of weak and strong pinning should be considered. In the case of \emph{weak pinning}, i.e. when the depinning current density $j_c$ is much less than the instability current density $j^{\ast}$, the linear flux-flow regime is realized for $j_{c}\ll j\ll j^{\ast}$ and the pinning effect on FFI should be negligibly small. In the opposite limiting case the absence of the CVC linearity just below the instability point may indicate that \emph{strong pinning} is essential for determining the FFI parameters. The case of strong pinning is, in particular, interesting for the use of nanopatterned superconductors in fluxonic applications \cite{Dob15apl,Dob15met}.

The objective of this paper is to provide a theoretical account for the pinning effect on the FFI critical parameters, namely the current density $j^{\ast}$, the electric field $E^{\ast}$, the resistivity $\rho^{\ast}$, the power density $P^{\ast}$, and the vortex velocity $v^{\ast}$. The problem is considered within the framework of the LO model and its BS generalization at the substrate temperature $T_{0}$ close to $T_{c}$ of a nanostructured superconducting film. The treatment of the problem is at once based on the BS approach, because it contains the LO results in the natural limiting case $B\ll B_{T}$.

The paper is organized as follows. In Sec. \ref{sProblem} the phenomenological BS approach is extended to account for pinning effects on the flux-flow instability parameters. To model the pinning, the simplest form of a cosine washboard pinning potential is used. The rigorous results of Sec. \ref{sCritPar} are analyzed in the limit of weak pinning in Sec. \ref{sWeakPin} and presented graphically in Sec. \ref{sAnalysis} in a broad range of magnetic field values and pinning strengths. A general discussion of the obtained results concludes our presentation in Sec. \ref{sDiscussion}.

\section{Main results}

\subsection{Formulation of the problem}
\label{sProblem}

\begin{figure}
\centering
    \includegraphics[width=1\linewidth]{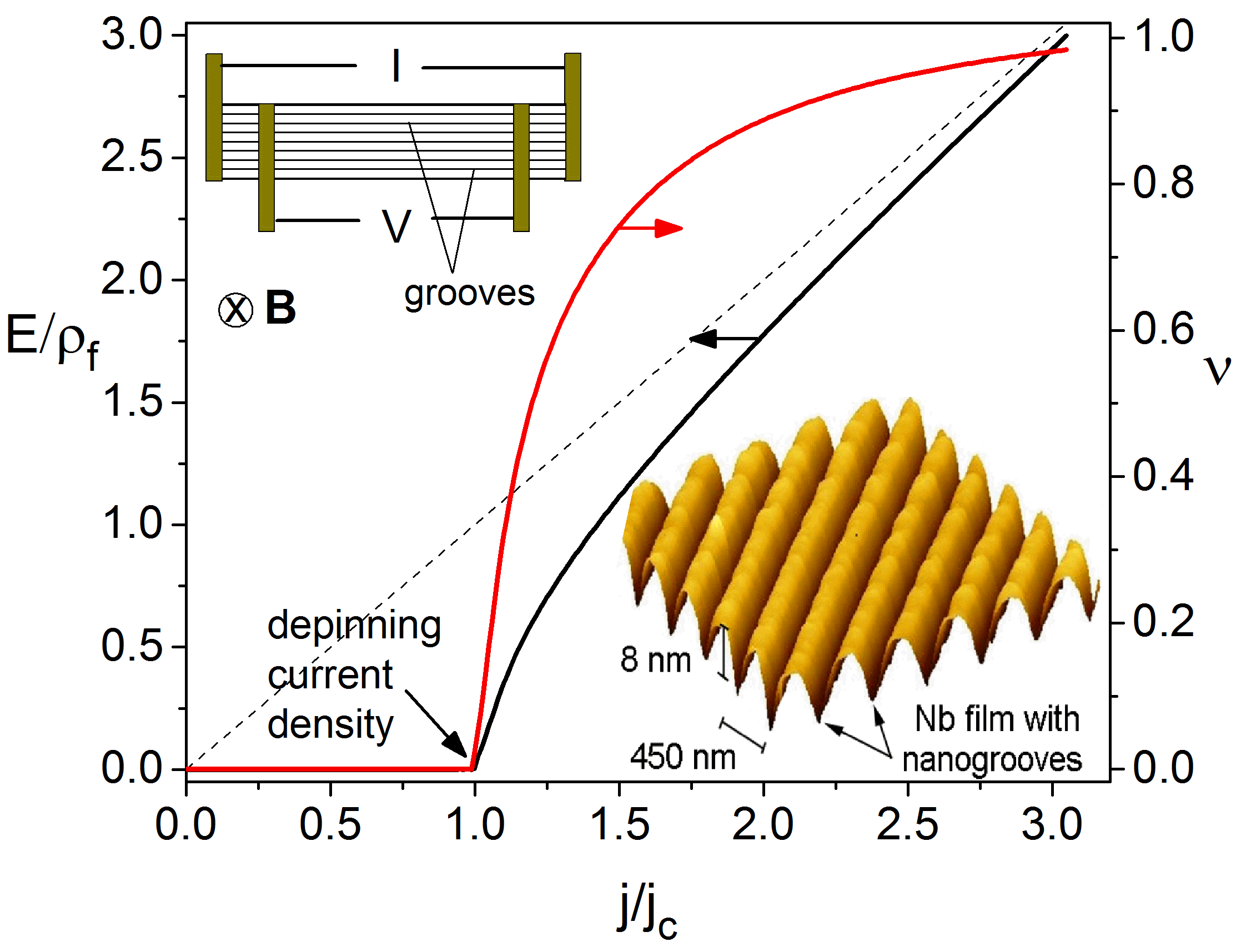}
    \caption{Left axis: The nonlinear current-voltage curve (CVC) $E(j)$ calculated for a cosine washboard pinning potential (WPP) of Ref. \cite{Shk08prb} in the limit of low temperatures. The dashed line corresponds to the free flux-flow regime $E/\rho_f = j$, where $\rho_f$ is the flux-flow resistivity. Right axis: The respective nonlinear function $\nu(j)$ calculated by Eq. (27) of Ref. \cite{Shk08prb}. Upper inset: Experimental geometry. The transport current $I$ is applied along the WPP channels (grooves). The vortex dynamics across the grooves leads to the appearance of the voltage $V$. Lower inset: Atomic force microscope image of a Nb film surface with a nanogroove array milled by focused ion beam \cite{Dob16sst} and inducing a pinning potential of the washboard type.}
   \label{fig1}
\end{figure}

For simplicity, a geometry is considered when the transport current is applied along the channels of a washboard pinning potential (WPP), see the upper inset in Fig. \ref{fig1}. In this case the nonlinear CVC of the sample can be presented as
\begin{equation}
    \label{1.1}
    \sigma E=j\nu\left(j\right),
\end{equation}
where $E$ is the longitudinal electric field and $j$ is the transport current density. Here, $\nu=\nu\left(j,T\right)$ is a nonlinear function which can be considered as the $(j,T)$-dependent effective mobility of the vortex under the action of the dimensionless driving Lorentz force $j$. Since $0<\nu<1$, this function can be treated as the probability of vortex hopping over the titled WPP barrier \cite{Shk06prb,Shk08prb}. At $T=0$, $\nu(j)$ is a steplike function with the condition $\nu(j)=0$ for $j<j_{c}$, where $j_{c}$ is the depinning current density. For simplicity, only this regime for $\nu(j)$ will be considered in what follows. If $j_{c}=0$, then $\nu(j)=1$ and the linear CVC $\sigma E=j$ follows from Eq. (\ref{1.1}) for the Bardeen-Stephen \cite{Bar65prv} conductivity $\sigma$.

For nanostructured superconductors, where the vortex dynamics can be described as their motion in a cosine WPP \cite{Mar76prl,Cof91prl,Che91prb,Maw97prb,Shk06prb,Shk08prb,Shk11prb,Dob12njp,Gui14nph,Dob15apl,Dob15met,Dob16sst}, the nonlinear CVC of the sample can be presented as \cite{Shk08prb}
\begin{equation}
    \label{7}
    \sigma E=\sqrt{j^{2}-j_{c}^{2}}\qquad\mathrm{or}\qquad
    j=\sqrt{j_{c}^{2}+\sigma^{2} E^{2}},
\end{equation}
where $j_{c}$ is the critical (depinning) current density indicated in Fig. \ref{fig1}.
Here
\begin{equation}
    \label{8}
    \sigma =\sigma_{f}(1-T/T_{c})^{-1/2}[1+(E/E^{\ast}_{LO}(T))^{2}]^{-1}f(B/H_{c2})
\end{equation}
obeys the LO expression for the nonlinear $E$-dependent flux-flow conductivity near $T_c$ in the dirty limit \cite{Lar86inb}. The function $f(B/H_{c2})$ in Eq. (\ref{8}), which takes into account the vortex core overlap, was tabulated in \cite{Lar86inb}. For magnetic field values of interest $B\leq 0.4H_{c2}(T)$, it amounts to $f(B/H_{c2})\simeq 4.04$ \cite{Bez92pcs}. With $E^{\ast}=v^{\ast}(B/c)$, where the critical velocity $v^{\ast}$ is given by Eq. (\ref{4}), for $E^{\ast}_{LO}(T)$ in Eq. (\ref{8}) it was shown \cite{Lar75etp} that
\begin{equation}
    \label{9}
    [E^{\ast}_{LO}(T)]^{2}=(B/c)^{2}(D/\tau_{\varepsilon})(\sqrt{14\zeta(3)}/\pi)(1-T/T_{c})^{1/2}.
\end{equation}
In Eq. (\ref{9}) $\zeta(3)=1.202$ is the Riemann zeta function of 3. For the upper critical field near $T_{c}$ applies
\begin{equation}
    \label{10}
    H_{c2}(T)=(4\Phi_{0}/\pi^{2}\hbar D)k_B (T_{c}-T),
\end{equation}
which is valid for superconductors with a short electron mean free path. Here $\Phi_{0}=\pi\hbar c/e_{0}$ is the magnetic flux quantum, $e_{0}$ is the electron charge, and $k_B$ is the Boltzman constant.

Following the BS approach we note that the quasiparticles temperature in Eq. (\ref{8}) is independent of the electric field $E$. It should be found from the heat balance equation (\ref{5}), which in the presence of pinning reads
\begin{equation}
    \label{11}
    h(T_{0})(T-T_{0})=dE\sqrt{j_{c}^{2}+\sigma^{2} E^{2}}.
\end{equation}

For the solution of Eq. (\ref{11}) in conjunction with Eq. (\ref{6}) we introduce, following Ref. \cite{Bez92pcs}, new dimensionless variables  $e\equiv E^{\ast}/E^{\ast}_{LO}(T_0)$ and $t\equiv (T_{c}-T^{\ast})/(T_{c}-T_{0})$. In addition, we take into account that Eq. (\ref{8}) can be identically transformed, using Eq. (\ref{10}), to
\begin{scriptsize}
\begin{equation}
\begin{array}{lll}
    \label{12}
    \sigma (E,T,T_{0})=
    \\
    =
    \displaystyle\frac{16.16  \sigma_{n}c k_B(T_{c}-T)}{\pi e_{0}DB(1-T_{0}/T_{c})^{1/2}[\sqrt{(T_{c}-T)/(T_{c}-T_{0})}+E^{2}/E^{2}_{LO}(T_{0})]}.
\end{array}
\end{equation}
\end{scriptsize}

Now one can show that in the presence of pinning, the BS extremum condition given by Eq. (\ref{6}) leads to the same result as in the absence of pinning (Eq. (29) in \cite{Bez92pcs})
\begin{equation}
    \label{13}
    1+\frac{e}{2t}\frac{dt}{de}-\frac{e^{2}}{\sqrt{t}}\left(1-\frac{e}{t}\frac{dt}{de}\right)=0.
\end{equation}
Note that Eq. (\ref{13}) derived in the presence of pinning does not depend on $j_{c}$ explicitly. From the heat balance equation (\ref{11}) it is possible to derive $dt/de$ and eliminate it from Eq. (\ref{13}). To accomplish this, one first finds the derivative of Eq. (\ref{11}) with respect to $E$ and evaluates it at the critical point given by Eq. (\ref{6}). The result is
\begin{equation}
    \label{14}
    dt/de=-(b/j_{0})\sqrt{j^{2}_{c}+j^{*2}_{0}}.
\end{equation}
Here $b=B/B_{T}$, $j^{\ast}_{0}=j_{0}[2et/(e^2+\sqrt{t})]$, and as in \cite{Bez92pcs},
\begin{equation}
\begin{array}{lll}
    \label{15}
    B_{T}=0.298hc\tau_{\varepsilon}e_{0}k_{B}^{-1}R_{\Box},
    \\[2mm]
    j_{0}=2.91(\sigma_{n}/e_{0})(D\tau_{\varepsilon})^{-1/2}(k_{B}T_{c})(1-T_{0}/T_{c})^{3/4}.
\end{array}
\end{equation}
The parameter $R_{\Box}=(\sigma_{n}d)^{-1}$ is the sheet resistance. Next, in the dimensionless form Eq. (\ref{11}) reads
\begin{equation}
    \label{16}
    1-t=-(be/j_{0})\sqrt{j^{2}_{c}+j^{*2}_{0}}.
\end{equation}
Then, from Eqs. (\ref{14}) and (\ref{16}) follows
\begin{equation}
    \label{17}
    dt/de=-(1-t)/e.
\end{equation}
Finally, the elimination of $dt/de$ from Eqs. (\ref{13}) and (\ref{17}) yields
\begin{equation}
    \label{18}
    e^{2}=\sqrt{t}(3t-1)/2.
\end{equation}
It is worth noting that Eq. (\ref{18}), which relates the variables $e$ and $t$, coincides with Eq. (32) in \cite{Bez92pcs}, that is pinning has no influence on this relation.

Now, taking into account that $j_{0}^{\ast}=j_{0}[2et/(e^2+\sqrt{t})]$, it is possible to derive from Eqs. (\ref{14}), (\ref{17}), and (\ref{18}) an equation for the dependence $t=t(b,\alpha)$, where $\alpha\equiv j_{c}/j_{0}$,
\begin{equation}
    \label{19}
    \frac{2(1-t)^2}{b^2\sqrt{t}(3t-1)}=\alpha^{2}+\frac{8t\sqrt{t}(3t-1)}{(3t+1)^{2}}.
\end{equation}
For $\alpha=0$ (i.e. in the absence of pinning, when $j_{c}=0$) Eq. (\ref{19}) has the solution $t=t(b)$ obtained previously by BS \cite{Bez92pcs}:
\begin{equation}
    \label{20}
    t=[1+b+(b^2+8b+4)^{1/2}]/3(1+2b).
\end{equation}
That is, quasiparticles overheating is $b$-dependent and it is given by Eq. (\ref{20}). The equation for $t(b)$ can be rewritten as
\begin{equation}
    \label{21}
    b(t)=(1-t)(3t+1)/2t(3t-1).
\end{equation}
Since $t>0$ per definition, $1/3 < t(b) <1$ follows from Eq. (\ref{21}). From Eq. (\ref{21}) it also follows that $dt/db<0$ and $t(b)$ monotonically decreases with $t(b=1)\simeq 2/3$. The LO limit ensues at $t\rightarrow 1$ and $b\rightarrow 0$, while the BS limit corresponds to $t\rightarrow 1/3$ and $b\rightarrow \infty$.

\subsection{Pinning effects on FFI parameters}
\label{sCritPar}

Equation (\ref{19}) is the key equation for the subsequent analysis. When $\alpha\neq 0$, the solutions of Eq. (\ref{19}) yield the dependences $t(\alpha,b)$ which will be used for the calculation of the FFI critical parameters in the presence of a cosine WPP. Namely, these are the critical electric field $E^{\ast}$, the critical velocity $v^{\ast}$, the critical current density $j^{\ast}$, the critical power $P^{\ast}=E^{\ast}j^{\ast}$, and the critical conductivity $\sigma^{\ast}(t)=j^{\ast}/E^{\ast}$.

In the presence of pinning, Eq. (\ref{19}) yields  $t=t(\alpha,b)$ and $t(\alpha,b)\rightarrow t(b)$ at $\alpha\rightarrow 0$. With the definition of $t$ one obtains
\begin{equation}
    \label{22}
    T^{\ast}(t)=T_{0}+(1-t)( T_{c}-T_{0}),
\end{equation}
where $T^*(t)$ is a linearly decreasing function of $t=t(\alpha,b)$ for $1/3<t<1$.

Proceeding to the electric field, we note that since $e=E^{\ast}/E^{\ast}_{LO}(T_{0})$, one has $E^{\ast}(t)=e(t)E^{\ast}_{LO}(T_{0})$. Then using Eq. (\ref{9}) one obtains $E^{\ast}_{LO}(T_{0})=bE_{0}$, where
\begin{equation}
    \label{222}
    E_{0}=1.143(B_T/c)(D/\tau_{\varepsilon})^{1/2}(1-T_{0}/T_{c})^{1/4},
\end{equation}
and the physical meaning of $E_{0}$ was discussed in the BS work (see Eq. (34) in \cite{Bez92pcs}). Finally, using Eq. (\ref{18}), one derives
\begin{equation}
    \label{23}
    E^{\ast}(t)/E_{0}=be(t)=bt^{1/4}\sqrt{3t-1}/\sqrt{2}.
\end{equation}

With the relation $v^{\ast}=(c/B)E^{\ast}$ for the critical velocity and Eq. (\ref{23}) one obtains that $v^{\ast}(t)$ in the presence of pinning is
given by
\begin{equation}
    \label{24}
    v^{\ast}(t)=e(t)v^{\ast}_{LO}.
\end{equation}
Here, $v^{\ast}_{LO}$ is the LO critical velocity independent of $E$ and $B$. It is given by Eq. (\ref{4}) at $T=T_0$. From Eq. (\ref{24}) it is clear that the dependence of $v^{\ast}(t)$ on $\alpha$ and $b$ is mediated by the dependence $e(t)$ through Eq. (\ref{18}).

We proceed now to an analysis of the dependence $j^{\ast}(t)$. From Eq. (\ref{8}) taken at the critical point we obtain
\begin{equation}
    \label{25}
    j^{\ast}(t)=\sqrt{j_{c}^{2}+(\sigma^{\ast}E^{\ast})^{2}},
\end{equation}
where, as follows from Eqs. (\ref{12}), (\ref{18}) and the definitions of $e$ and $t$,
\begin{equation}
    \label{26}
    \sigma^{\ast}E^{\ast}=j_0\cdot 2et/(e^2+\sqrt{t}).
\end{equation}
Then, using Eq. (\ref{21}), Eq. (\ref{25}) can be transformed to
\begin{equation}
    \label{27}
    j^{\ast}/j_0=\sqrt{\alpha^2+8t^{3/2}(3t-1)/(3t+1)^2},
\end{equation}
Finally, comparing  Eq. (\ref{27}) and Eq. (\ref{19}), where also $j_0^{\ast}=j_0\cdot 2et/(e^2+\sqrt{t})$  and $j_0^{\ast}$ is equal to the right-hand side of Eq. (\ref{26}), it follows that
\begin{equation}
    \label{28}
    j^{\ast}/j_0=(1-t)/be.
\end{equation}
Now, having Eq. (\ref{23}) for $E^{\ast}(t)/E_0$ and Eq. (\ref{28}) for $j^{\ast}(t)/j_0$ it is clear that
\begin{equation}
    \label{29}
    P^{\ast}(t)/P_0=1-t,
\end{equation}
where
\begin{equation}
    \label{30}
    P_0=E_0 j_0=(h/\alpha)(T_{c}-T_{0}).
\end{equation}
In Eq. (\ref{29}) $P^{\ast}=E^{\ast}j^{\ast}$ is the critical power density dissipated in the film, while $P_0$ is the power density corresponding to the temperature difference $\Delta T_{co}=T_{c}-T_{0}>0$.

\begin{figure}
\centering
    \includegraphics[width=1\linewidth]{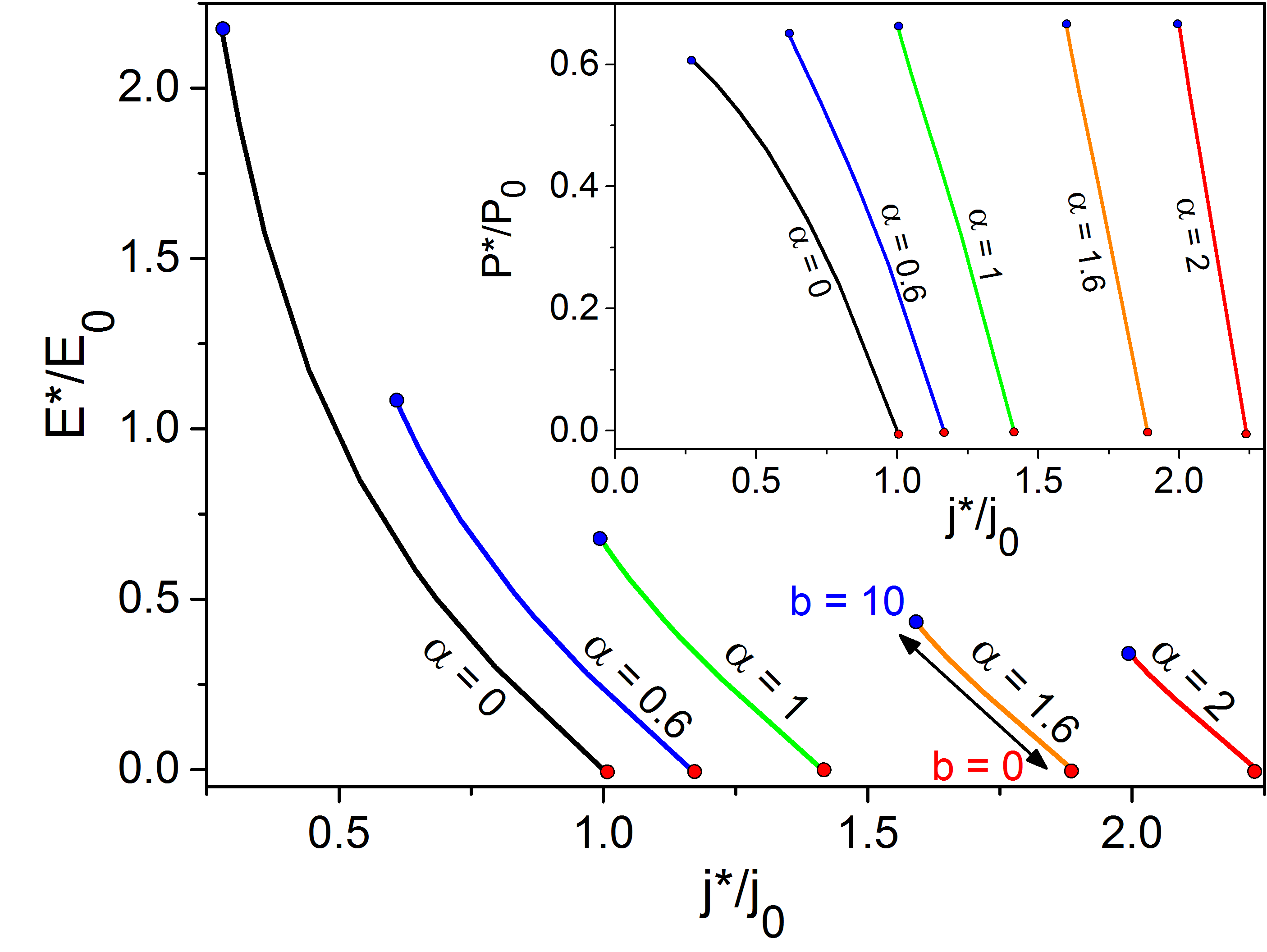}
    \caption{Normalized electric field $E^\ast/E_0$ versus normalized current density $j^\ast/j_0$ at the instability point for a series of values of the pinning strength parameter $\alpha \equiv j_c/j_0$. For all curves, the blue spots outline values obtained at $b\equiv B/B_T = 10$, while the red ones mark those at $b=0$. The curve for $\alpha =0$ coincides with the BS curve in Fig. 1 in \cite{Bez92pcs}. Inset: Normalized power density $P^\ast/P_0$ versus normalized current density $j^\ast/j_0$ at the instability point for the same series of values of the pinning strength parameter $\alpha$.}
   \label{fig2}
\end{figure}

Using Eqs. (\ref{23}) and (\ref{28}) it is possible to calculate the critical conductivity $\sigma^{\ast}(t)$ of the sample
\begin{equation}
    \label{32}
    \sigma^{\ast}(t)=j^*(t)/E^{\ast}(t)=\sigma_{0}(1-t)/(be)^2,
\end{equation}
where
\begin{small}
\begin{equation}
    \begin{array}{ll}
    \label{33}
\sigma_{0}\equiv j_{0}/E_{0}=
\\[1mm]
=
(8.08/\pi)^{2}(4,1/\pi)^{3/2}\sigma_{n}N(0)k_{B}(k_{B}T_c)\sqrt{1-T_{0}/T_{c}}/h\tau_{\varepsilon},
    \end{array}
\end{equation}
\end{small}
$\sigma_{n}=N(0)De_{0}^{2}$, and $N(0)=mp_{F}/\pi^{2}\hbar^{3}$ is the density of states. A series of $E^\ast/E_0$ versus $j^\ast/j_0$ curves at the instability point is plotted in Fig. \ref{fig2} for a series of values of the pinning strength parameter $\alpha = j_c/j_0$.

\subsection{Weak pinning}
\label{sWeakPin}

For convenience, we introduce the dimensionless pinning pinning strength parameter $\alpha\equiv j_{c}/j_{0}$, the dimensionless magnetic field $b\equiv B/B_{T}$, and denote
\begin{equation}
    \label{e1}
    x(t) \equiv \frac{2(1-t)^2}{ b^2\sqrt{t}(3t -1)},\qquad y(t) \equiv \frac{8t^{3/2}(3t-1)}{(3t +1 )^2}.
\end{equation}
In the case of weak pinning (i.e. $\alpha \rightarrow 0$), Eq. (\ref{19}) can be solved analytically using $\alpha \rightarrow 0$ as a small perturbation parameter. Thus, in the limiting case $\alpha^2 =0$ one obtains $x(t_0) = y(t_0)$, where $t_0 = t(b)$ was obtained in the BS work \cite{Bez92pcs} and it is given by Eq. (\ref{20}). Accordingly, for $\alpha^2\rightarrow0$ one can write $t(b,\alpha) \simeq t_0 - \varepsilon$, where $\varepsilon = A(b)\alpha^2 \ll 1$. Then it is possible to express the functions $x(t)$ and $y(t)$ at $t = t_0 - \varepsilon$ in terms of $x(t_0)$, $y(t_0)$, and $\varepsilon$, namely
\begin{equation}
\begin{array}{lll}
\displaystyle
    x(t_0 - \varepsilon) \simeq x(t_0)\left[1 + \varepsilon\left(\frac{2}{1-t_0} + \frac{1}{2t_0} + \frac{3}{3t_0 -1}\right)\right],\\[4mm]
\displaystyle
    y(t_0 - \varepsilon) \simeq y(t_0)\left[1 - \varepsilon\left(\frac{3}{2t_0} + \frac{3}{3t_0 -1} - \frac{6}{3t_0 +1}\right)\right].
\end{array}
\end{equation}

This leads to the following equation for $A(b)$
\begin{equation}
\label{e4}
    A(b) = \frac{t_0(1-t_0)(9t^2 -1)}{2x_0(3t^2_0 + 6t_0 -1)},
\end{equation}
which can be solved in the limiting cases of \emph{weak} ($b\rightarrow 0$) and \emph{strong} ($b\gg 1$) magnetic fields. Namely, when $b\rightarrow 0$, one obtains $t_0(b \rightarrow 0) \simeq 1- b$. Then, in the main approximation in $b \rightarrow 0$, from Eq. (\ref{e4}) it follows that $A(b\rightarrow 0) \simeq b/2$ and
\begin{equation}
\label{e6}
    t(b\rightarrow0) \simeq 1 - b(1 + \alpha^2/2).
\end{equation}

When $b\gg1$, from Eq. (\ref{19}) one obtains $t_0(b\gg 1) \simeq(1 + 2/b)/3$. Equation (\ref{e4}) yields $A(b\gg1) \simeq 2/\sqrt{3}$ and
\begin{equation}
\label{e7}
    t(b\gg1) \simeq (1 + 2/b)/3 - 2\sqrt{3}\alpha^2.
\end{equation}

\section{Graphical analysis}
\label{sAnalysis}
\begin{figure*}
\centering
    \includegraphics[width=0.44\linewidth]{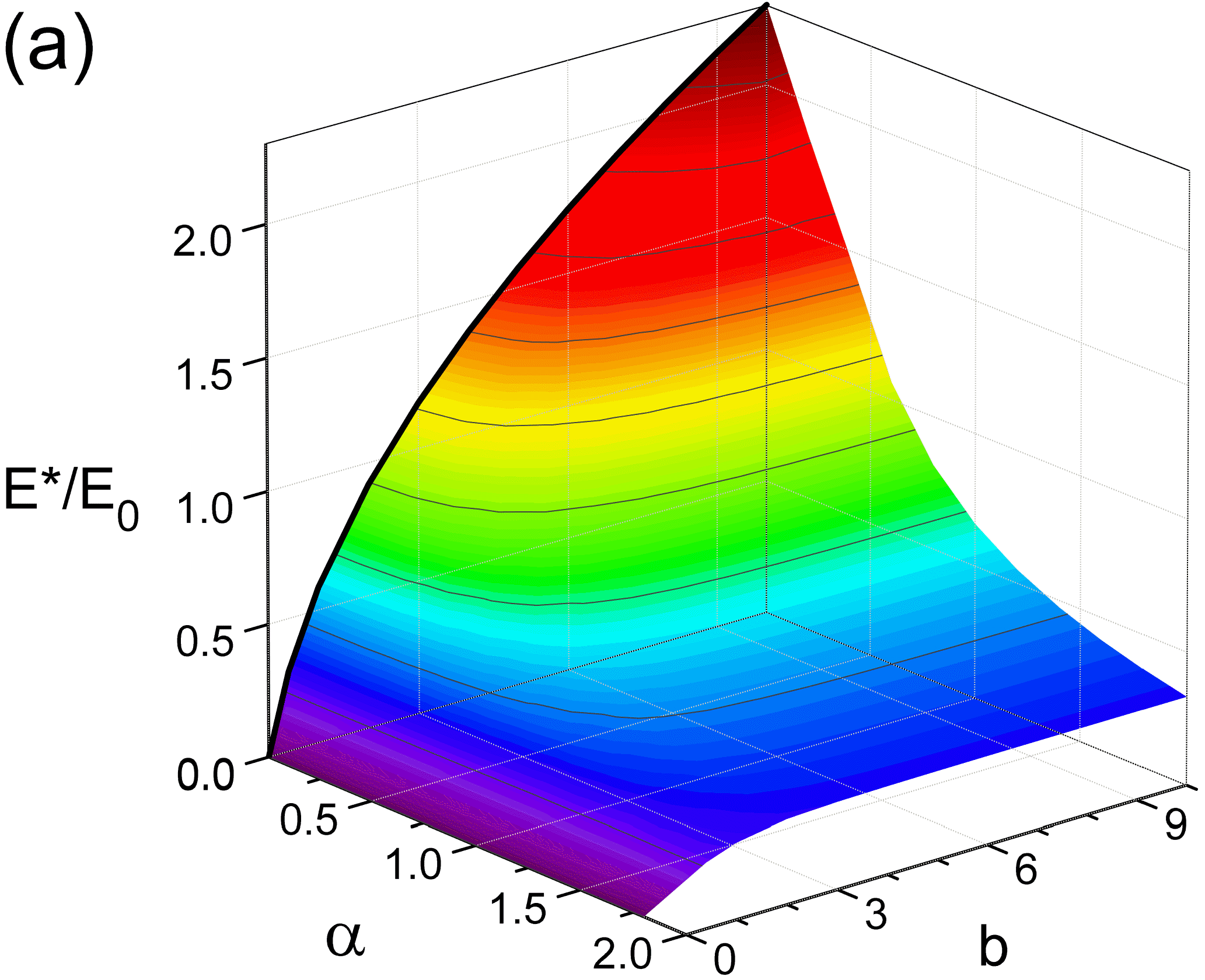}\hspace{5mm}
    \includegraphics[width=0.44\linewidth]{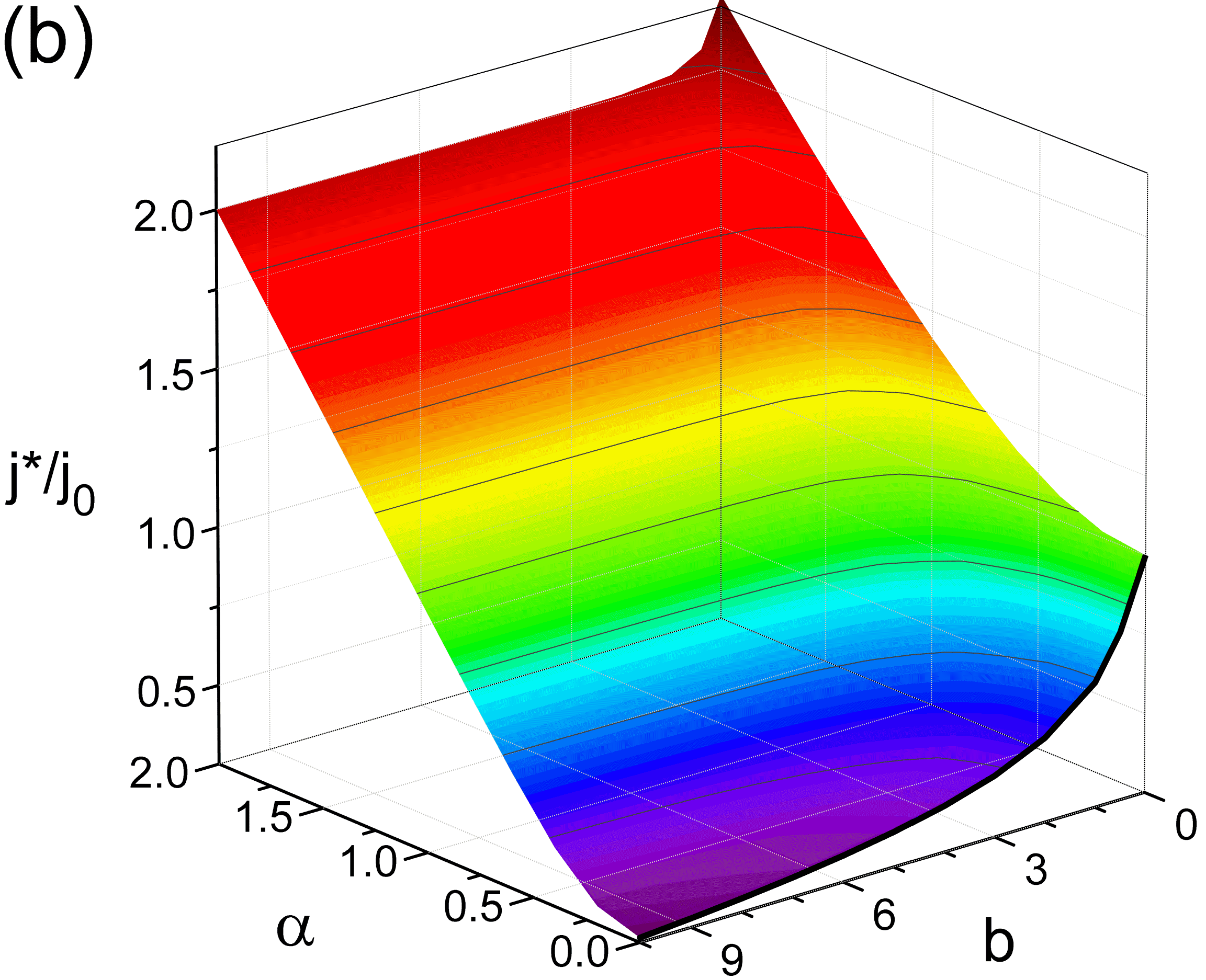}\vspace{3mm}
    \includegraphics[width=0.44\linewidth]{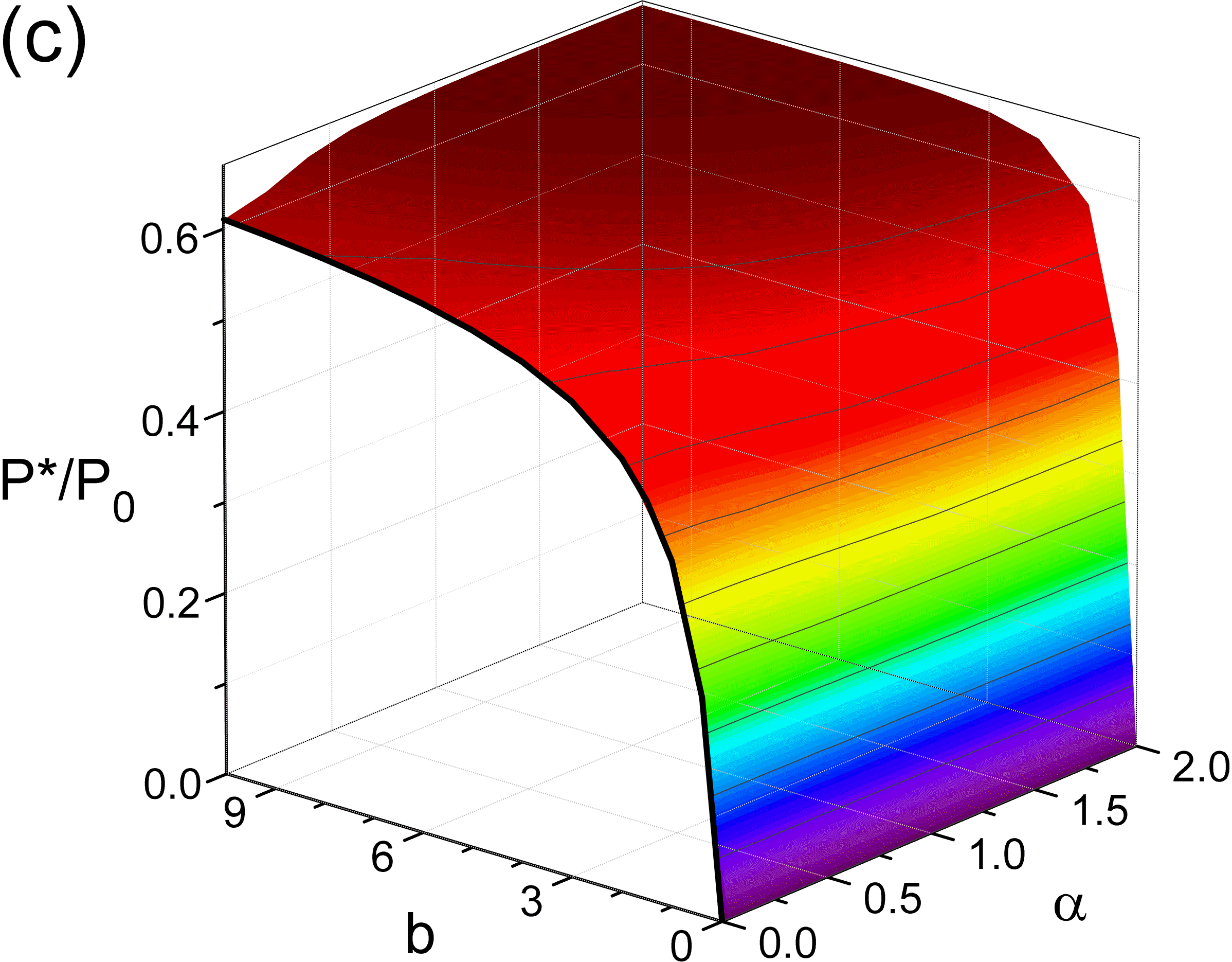}\hspace{5mm}
    \includegraphics[width=0.44\linewidth]{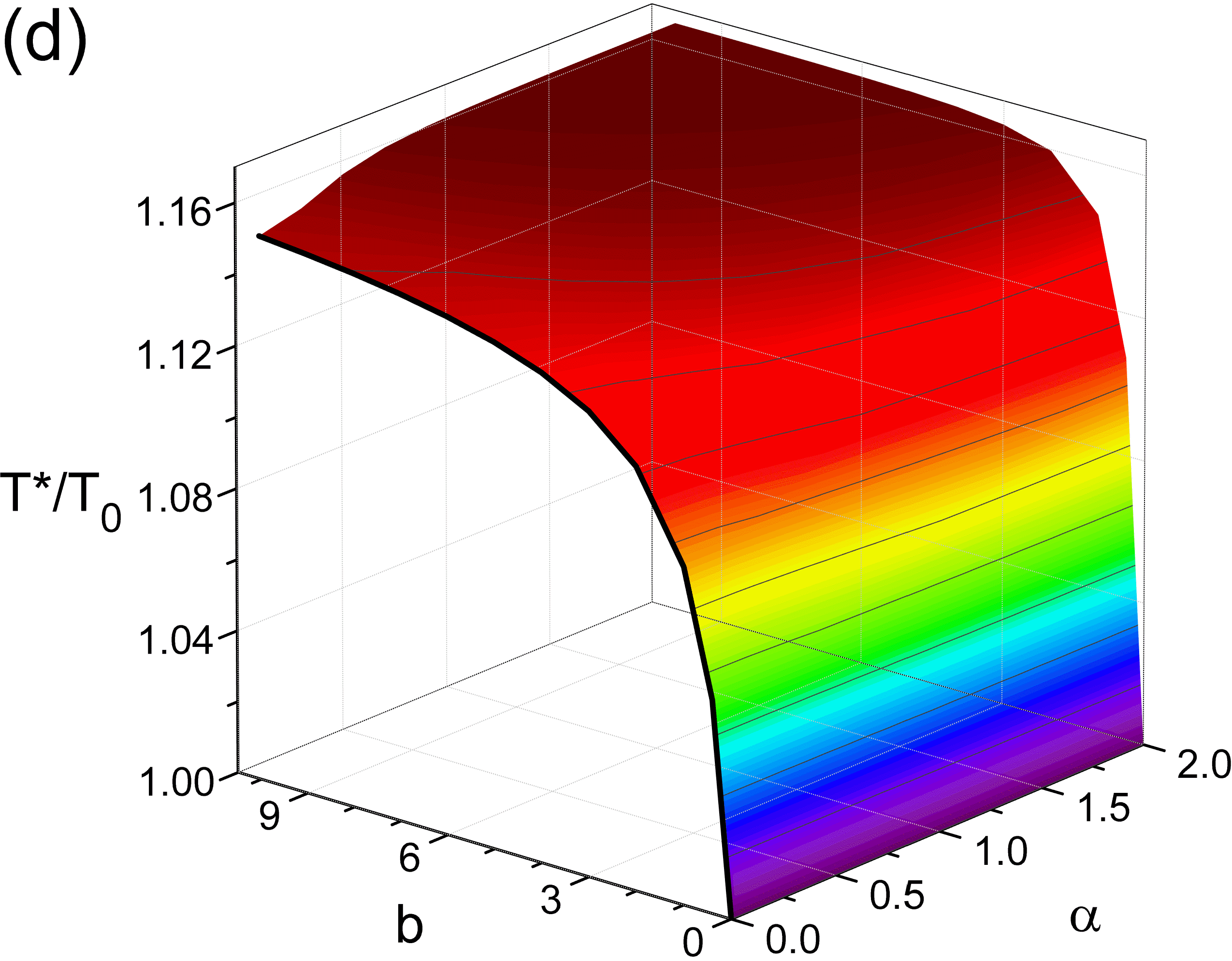}
    \caption{Dependence of the normalized critical electric field $E^{\ast}(t)/E_{0}$ (a), critical current density $j^{\ast}(t)/j_{0}$ (b), critical power density $P^{\ast}(t)/P_{0}$ (c), and critical temperature $T^{\ast}(t)/T_{0}$ (d) on the dimensionless pinning strength parameter $\alpha \equiv j_c/j_0$ and the dimensionless magnetic field $b = B/B_T$ calculated by Eqs. (\ref{23}), (\ref{27}), (\ref{29}), and (\ref{22}), respectively. The parameters $B_T$ and $E_{0}$ are given by Eqs. (\ref{15}) and (\ref{222}). The thick black curves at $\alpha = 0$ reproduce the BS results \cite{Bez92pcs} without pinning.}
   \label{figs}
\end{figure*}
The objective of this section is to graphically analyze the dimensionless FFI critical parameters in the presence of pinning as functions of the dimensionless pinning strength parameter $\alpha\equiv j_{c}/j_{0}$ and the dimensionless magnetic field $b\equiv B/B_{T}$, in a broad range of the respective parameters.

We begin our analysis with the critical electric field $E^{\ast}(t)/E_{0}$ calculated by Eq. (\ref{23}) and plotted in Fig. \ref{figs}(a). We reveal that at any arbitrary value of $\alpha$ ($0<\alpha<2$), $E^{\ast}(t)/E_{0}$ is a monotonically increasing function of $b$. As the first check, we have proven that in the limit of no pinning at $\alpha=0$ the dependence $E^{\ast}(t)/E_{0}$ coincides with the curve obtained by BS (Fig. 2 in \cite{Bez92pcs}). The dependence $E^{\ast}(t)/E_{0}$ is nonlinear, while with increasing $\alpha$ the nonlinearity is weakening as $E^{\ast}(t)/E_{0}$ tends to become independent of $b$ for $b \gtrsim 2$. In addition, one sees that whereas $E^{\ast}(t)/E_{0} = 0$ for all $\alpha$ at $b = 0$, at larger $b$ values $E^{\ast}(\alpha)/E_{0}$ becomes a rapidly decreasing function of the pinning strength.

In contrast to $E^{\ast}(t)/E_{0}$ in Fig. \ref{figs}(a), which monotonically \emph{decreases with increasing} $\alpha$, the behavior of $j^{\ast}(t)/j_0$ calculated by Eq. (\ref{27}) as a function of $\alpha$ at $b=const$ is opposite. Namely, $j^{\ast}(t)/j_0$ strongly \emph{increases} with $\alpha$, see Fig. \ref{figs}(b). We again check that at $\alpha=0$ the magnetic field dependence of $j^{\ast}(t)/j_0$ coincides with the curve obtained by BS (see Fig. 3 in \cite{Bez92pcs}), which is a monotonically decreasing function of $b$. At larger values of $\alpha \simeq 1$ a part of the graph appears in Fig. \ref{figs}(b), where the critical current density is independent of $b$. This part has a tendency to expand with a further increase of $\alpha$.

Figure \ref{figs}(c) displays the critical power $P^{\ast}(t)/P_{0}$ calculated by Eq. (\ref{29}) as a function of $\alpha$ and $b$. The function $P^{\ast}/P_{0}(\alpha,b )$ is strongly increasing with $b$ for $0 < b \lesssim 2$, exhibits a plateau at $2 \lesssim b < 10$ and large $\alpha$, and it is only very weakly increasing as a function of $\alpha$ at a given value of $b$.

In Fig. \ref{figs}(d) the critical temperature $T^{\ast}(t)/T_{0}$ computed by Eq. (\ref{22}) as a function of $\alpha$ and $b$ at $T_{0}/T_{c}= 0.8$ is presented. While in the whole range of magnetic fields the relative changes in $T^{\ast}/T_{0}$ do not exceed $16\%$ for a fixed $\alpha$ and these changes primarily ensue at $0 < b \lesssim 3$, $T^{\ast}/T_{0}$ is practically independent of $\alpha$.

\section{Discussion}
\label{sDiscussion}
Before a deeper discussion of the theoretical results obtained in this work in comparison with those presented in \cite{Lar75etp} and later in \cite{Bez92pcs}, we need to briefly recall the main theoretical FFI features at temperatures near $T_c$ in the absence of pinning obtained initially in those works \cite{Lar75etp,Bez92pcs}. According to LO \cite{Lar75etp,Per05prb}, FFI for dirty films leads to the shrinkage of the vortex cores with increasing vortex velocity up to its critical value $v^{\ast}$, refer to Eqs. (\ref{2}) and (\ref{3}). Then, for current-driven measurements the nonlinear resistive part of the CVC exhibits a jump-like voltage rise at $v = v^{\ast}$, when the Lorentz force equals to the greatest value of the viscous damping force for a vortex. In experiments \cite{Mus80etp,Kle85ltp}, however, several discrepancies from the LO theoretical results were revealed, as already mentioned in the introduction. Later on, BS have shown \cite{Bez92pcs} that these discrepancies may have a common cause, namely the overheating of phonons and quasiparticles in the film due to the dissipation during the vortex motion. The LO results were generalized by BS in a such way, that for $B\ll B_T$ the overheating of the quasiparticles is small and the LO theory is valid, whereas for $B\gg B_T$ the overheating of the quasiparticles is important and it allows one to explain the most of discrepancies from the LO theory discussed in \cite{Mus80etp,Kle85ltp,Vol92fnt}. It is essential that both, the LO and BS theoretical results were obtained \emph{without accounting for vortex pinning}.

In the previous sections, within the framework of the BS approach, the influence of pinning on the FFI critical parameters for the current flow along the WPP channels at the substrate temperature $T_{0}$ close to the critical temperature $T_{c}$ of the nanostructured superconducting film (see Fig. \ref{fig1}) has been theoretically analyzed. While the FFI critical parameters in the BS paper \cite{Bez92pcs} have already been calculated as $b$-dependent ($b\equiv B/B_{T}$), in this paper an additional $\alpha$-dependence ($\alpha=j_{c}/j_{0}$) appears through Eq. (\ref{19}), thereby introducing the \emph{variable pinning strength}. The main tasks of the theoretical analysis was then to solve the main equation (\ref{19}) with respect to $t(\alpha,b)$ and to reveal the influence of pinning on the FFI critical parameters, namely $E^{\ast}$, $j^{\ast}$, $v^{\ast}$, $\sigma^{\ast}$, and $P^{\ast}$. As a natural passage to the limit $\alpha=0$, the solutions $t(\alpha,b)$ have been checked to coincide with the well-known BS results $t(b)$. From the graphical analysis it is clear that, at a given magnetic field $b=const$, \emph{with increasing pinning strength}  $E^{\ast}$ \emph{decreases}, $j^{\ast}$ \emph{increases}, while $P^{\ast}$ and $T^{\ast}$ are \emph{practically constant}.

We would like to stress that the introduction of pinning into the BS instability problem is phenomenological. Namely, instead of the linear CVC relation $j=\sigma (T)E$ (at $T=const$) with $\sigma (T)= \sigma_{n} H_{c2}(T)/B$ used in \cite{Bez92pcs} here a nonlinear CVC at $T=const$ generated by the cosine WPP and taken at $T=0$ has been used. The conductivity $\sigma(T)$ in Eg. (\ref{7}) is the same as that in Eq.~(\ref{1}). Being aware that the cosine WPP is a very particular form of the pinning potential, we emphasize that its advantage is a very simple CVC given by Eq. \eqref{7}. Obviously, for an arbitrary pinning potential the CVC can not be described by a simple analytical expression.

Theoretically, it is also possible to use the CVC derived for the cosine WPP at $T>0$ \cite{Shk08prb}, but here the case $T=0$ has been considered for simplicity. Further arguments in favor of the consideration of the CVC at $T =0$ are related to the two main features which are crucial for the considered problem: (i) The CVC at $T =0$ exhibits a nonlinear transition from the dissipation-free regime to the regime of viscous flux flow. (ii) The nonlinear transition at $T =0$ allows for a straightforward determination of the depinning current density $j_c$ corresponding to the onset of the vortex motion. Such a simple determination of $j_c$ at $T >0$ is impossible without introduction of further criteria. Obviously, the CVC at $T = 0$ is characterized by the parameter $j_c$ which, in turn, depends on the parameters of the WPP.

We now turn to suggestions for an experimental examination of the theoretical results obtained here. First of all, while a saw-tooth \cite{Shk99etp,Shk06prb,Luq07prb,Shk09prb} and a harmonic \cite{Mar76prl,Cof91prl,Che91prb,Maw97prb,Shk08prb,Shk11prb,Shk12inb,Shk14pcm} potential are the most simple particular WPP forms widely used in theoretical modeling, they are found across numerous experimental systems. These systems range from naturally occurring pinning sites in high-$T_c$ superconductors \cite{Ber97prl,Cha98sst,Pas99prl,Dan00prb} to artificially created linearly-extended pinning sites in superconductor thin films \cite{Nie69jap,Mor70prl,Yuz99pcs,Yuz99prb,Jaq02apl,Sil11sst,Hut02afm,Sor07prb,Dob12njp,Gui14nph,Dob16sst,Dob15apl,Dob15met,Dob15snm,Dob15mst,Dob10sst,Dob11snm,Dob11pcs}, see e.g. Ref. \cite{Dob17pcs} for a review. Systems with naturally occurring pinning sites are largely represented by cuprates in which one distinguishes the intrinsic pinning induced by the layered structure itself \cite{Ber97prl} and the planar pinning caused by uniaxial twins \cite{Cha98sst,Pas99prl,Dan00prb}. Artificial WPPs can be induced by a periodic thickness \cite{Nie69jap,Mor70prl} or magnetization \cite{Yuz99pcs,Yuz99prb,Jaq02apl,Sil11sst} modulation, thin film deposition onto facetted substrates \cite{Hut02afm,Sor07prb}, milling of periodically arranged nanogrooves in films by focused ion beam \cite{Dob12njp,Dob16sst,Dob15apl,Dob15met,Dob15snm,Dob15mst}, and decoration of films with ferromagnetic nanostripes by focused electron beam induced deposition (FEBID) \cite{Dob10sst,Dob11snm,Dob11pcs}.

A further feature of the CVC given by Eq.\,\eqref{7} is that it is derived within the framework of a \emph{single-vortex} model of anisotropic pinning \cite{Shk08prb}. For the most direct comparison of theory with experiment on, e.g. nanopatterned superconductors, this means that one has to tune the vortex dynamics in a coherent regime, when the entire vortex ensemble behaves as a vortex crystal. The background pinning due to undesired random disorder must be weak to ensure the long-range order correlations of the vortex lattice in the vicinity of the depinning transition. This can be realized e.g. in weak-pinning amorphous Mo$_3$Ge \cite{Gri15prb}, Nb$_{0.7}$Ge$_{0.3}$ \cite{Bab04prb}, and Al \cite{Sil12njp} films as well as in epitaxial thin films in the clean superconducting limit such as epitaxial Nb films on sapphire substrates \cite{Dob12tsf}.

The next experimental requirement is to perform measurements at so-called matching fields, when the Abrikosov vortex lattice is commensurate with the pinning landscape. That said, for a WPP each row of vortices shall be pinned at the position of linearly-extended pinning sites (for instance, nanogrooves shown in the lower inset to Fig. \ref{fig1}) and there will be neither vacant nanogrooves nor vortices pinned between them. It has been shown by computer simulations \cite{Luq07prb} that at the fundamental matching the vortex lattice has a crystalline structure, the effective vortex interaction is cancelled and the response of the vortex ensemble can be analyzed on the basis of that for a single vortex. The motion of vortices in this case is coherent, as can be seen, e.g. via ac/dc interference (Shapiro steps) in the CVCs \cite{Mar76prl,Dob15snm,Dob15mst}. These interference steps arise when the vortex hopping distance during one half ac cycle coincides with one or a multiple of the nanostructure period. Contrary, when $B$ is tuned away from the matching condition, the vortex lattice becomes jammed and the vortex motion lacks coherence. This means that for a particular superconductor with a given periodicity of the WPP, one should perform measurements at a fixed matching field value instead of probing the whole range of magnetic fields $0<B\lesssim 0.4B_{c2}$ where FFI is observed \cite{Bez92pcs}. At the same time, also the pinning strength parameter $\alpha$ is usually fixed for a given sample, with exception of cases \cite{Yuz99pcs,Yuz99prb,Sil12njp} where the pinning strength can be tuned by a proper magnetic state of the individual element. In this way, to systematically compare theory with experiment in a broad range of magnetic fields and pinning intensities, a series of samples with different nanostructure periods and pinning strengths is required. The use of WPP landscapes with a tunable pinning strength appears as an alternative promising approach. This should be possible by decoration of films with ferromagnetic nanostripes using FEBID \cite{Dob11snm,Dob10sst,Dob11pcs} in combination with post-growth processing \cite{Beg15nan,Dob15bjn} of samples for switching the magnetic state of the nanostripes and thereby changing the pinning strength.

The LO instability in superconducting films with a WPP has not been studied experimentally so far. For this reason, to examine our theoretical predictions, we would like to compare the main results of our phenomenological approach with an existing experiment \cite{Sil12njp} qualitatively. Silhanek \emph{et al} \cite{Sil12njp} investigated the LO instability by measuring the CVCs of a 50\,nm thick Al superconducting film deposited on top of an array of Py square rings. Their magnetic templates represent a flexible way to change the pinning strength by changing the magnetic state of the rings. Individual magnetic domains along each leg of the square rings were arranged to form either the so-called vortex state with a weak stray field or an onion state with a strong stray field. The experiment was carried out at $T/T_c = 0.89$ in the magnetic field range $0 < B < B_1 \ll B_T \approx 20$\,mT, where $B_1 = 1.808$\,mT is the matching field corresponding to one vortex per unit cell. That said, the experiment \cite{Sil12njp} was carried out in the low-field range, where overheating effects are not relevant. Importantly, at a fixed magnetic field value, the instability velocity $v^\ast$ has been revealed to decrease with increasing pinning strength, whereas the instability current density $j^\ast$ (slightly) decreased as the pinning strength increased, we refer to Fig. 5 in Ref. \cite{Sil12njp}.
\begin{figure}
\centering
    \includegraphics[height=6cm]{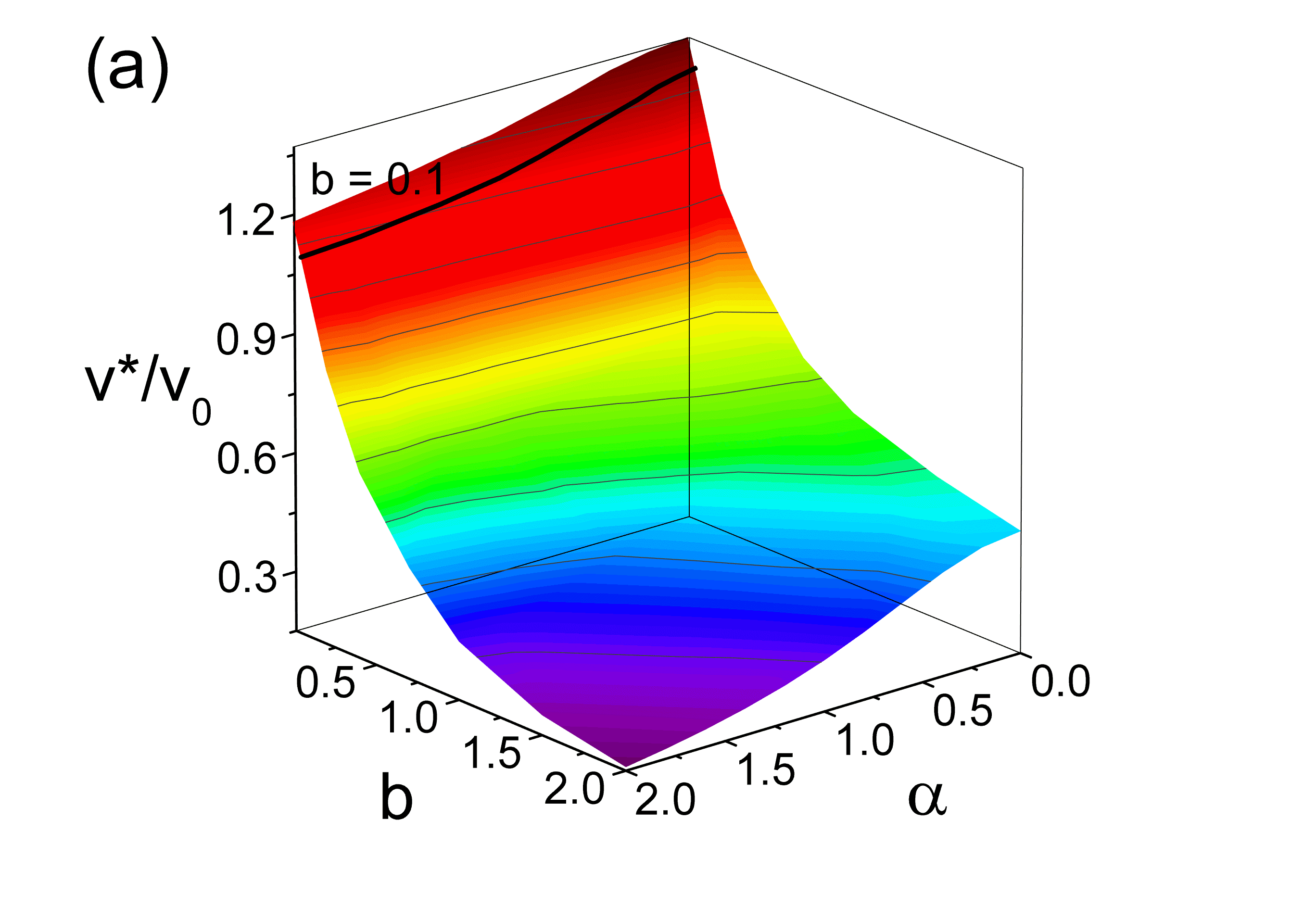}\vspace{2mm}
    \includegraphics[height=6cm]{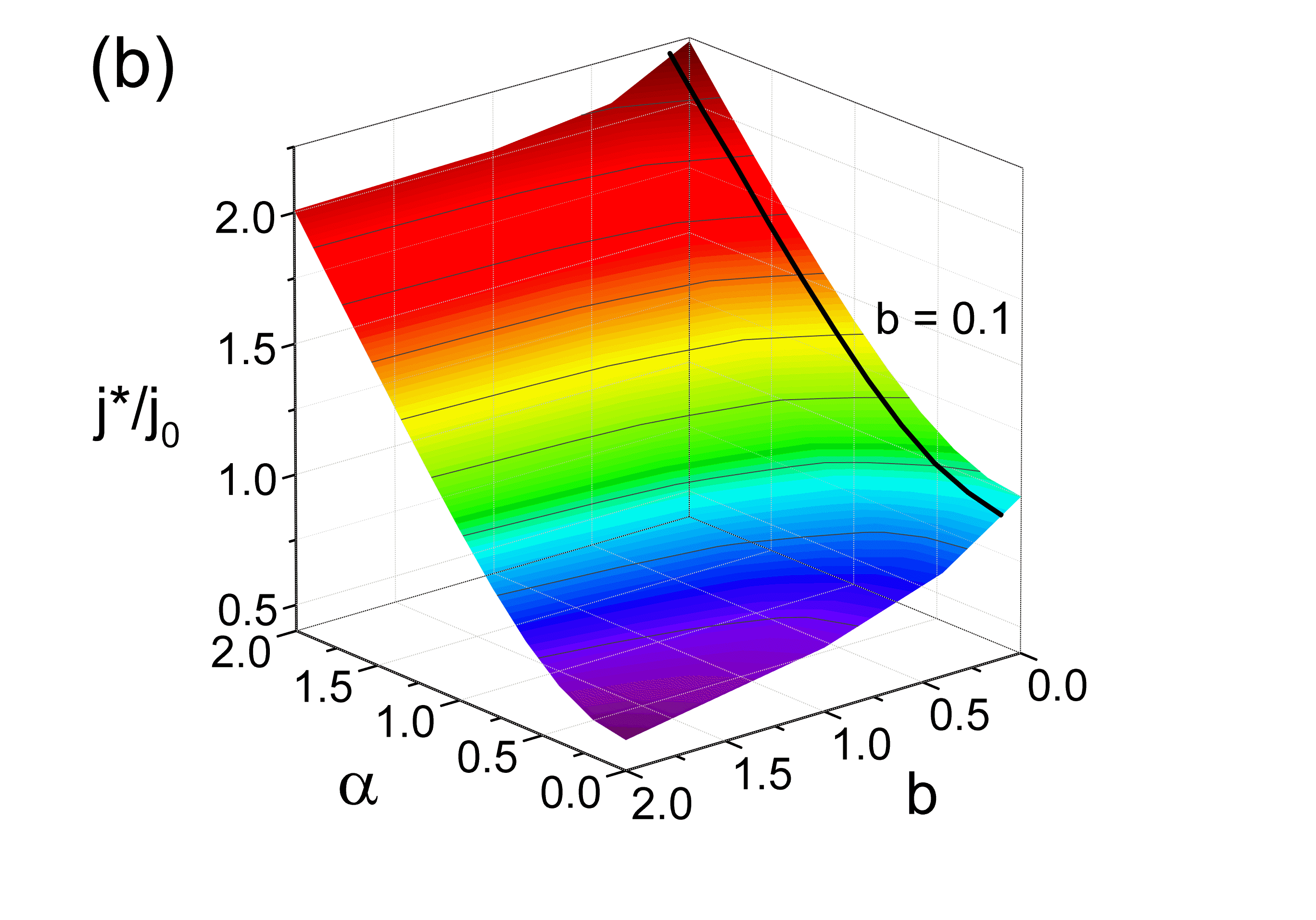}
    \caption{Dependence of the normalized instability velocity $v^{\ast}/v_{0}$ (a) and the instability current density $j^{\ast}/j_{0}$ (b) on the dimensionless pinning strength parameter $\alpha \equiv j_c/j_0$ and the dimensionless magnetic field $b = B/B_T$ calculated by Eqs. (\ref{23}) and (\ref{27}), respectively, in the low-field range. The parameters $B_T$ and $E_{0}$ are given by Eqs. (\ref{15}) and (\ref{222}) while $v_0 \equiv cE_0/B_T$. The thick black lines correspond to $b = 0.1$ for a comparison with the most closely related experiment of Silhanek \emph{et al} \cite{Sil12njp}, as detailed in the text.}
   \label{figZoom}
\end{figure}

To augment their experimental observations, Silhanek \emph{et al} \cite{Sil12njp} performed computer simulations relying upon the time-dependent Ginzburg-Landau equation. Their simulation results reported in Fig. 2 in Ref. \cite{Sil12njp} support both, the experimental observations of Ref. \cite{Sil12njp} and our theoretical predictions, as clearly seen in Fig. \ref{figZoom}, where $v^\ast(\alpha)$ and $j^\ast(\alpha)$ are marked by thick black lines at $b = 0.1 \approx B/B_T$ with $B$ being in the vicinity of their matching field $B_1$. In this way, even though the available experiment \cite{Sil12njp} was performed for a superconducting film with a different pinning potential, the main predictions of our phenomenological theory, namely \emph{a reduction of the instability velocity $v^\ast$ and an increase of the instability current density $j^\ast$ with increasing pinning strength} qualitatively agree with the results of both, electrical resistance measurements and computer simulations \cite{Sil12njp}. A systematic comparison of the experimental dependences $v^\ast(\alpha,b)$ and $j^\ast(\alpha,b)$ measured near $T_c$ with Eqs. (\ref{15}) and (\ref{222}) should include a broader range of the parameters $\alpha$ and $b$.

Finally, the introduction of pinning into the FFI problem in the opposite limiting case $T \ll T_c$ should be commented. At $T \ll T_c$ FFI is caused not by the standard LO scenario assuming a vortex shrinkage due to quasiparticles escaping from the vortex cores, but rather by the Kunchur hot-electron mechanism \cite{Kun02prl}: At $T \ll T_c$, when the electron-electron scattering time is shorter than the electron-phonon scattering time, $\tau_{ee} < \tau_{eph}$, the distribution function remains thermal-like and the electronic system exhibits a temperature rise with respect to the lattice. In consequence of this, additional quasiparticles are created thus leading to a diminishing of the superconducting order parameter $\Delta$. This results in an vortex expansion and a reduction of the viscous drag because of a softening of gradients of the vortex profile. All experimental observables were calculated in Ref. \cite{Kun02prl} as functions of the magnetic field value, by \emph{numerical} integration of the heat balance equation. The experimental results for YBCO were successfully fitted to the predicted $B$-dependences and the $j(E)$ curves \emph{in the absence} of pinning without any adjustable parameters \cite{Kun02prl,Kni06prb}.

The effect of pinning on the hot-electron FFI parameters has recently been analyzed theoretically in Ref. \cite{Shk17arx}. There, as in this work, the pinning is introduced phenomenologically by using the nonlinear conductivity generated by the WPP instead of the Bardeen-Stephen flux-flow conductivity in the CVC. A simpler heat balance equation for electrons in low-$T_c$ superconducting films has been solved in Ref. \cite{Shk17arx} in the two-fluid approach, \emph{without numerical integration} of the heat balance equation. A theoretical analysis has revealed \cite{Shk17arx} that the $B$-behavior of $E^\ast$, $j^\ast$ and $\rho^\ast$ is monotonic, whereas the $B$-dependence of $v^\ast$ is quite different as $dv^\ast/dB$ may \emph{change its sign twice}, as sometimes observed in experiments \cite{Leo10pcs,Gri12apl,Sil12njp,Gri09pcm,Gri10prb,Gri11snm}. The generalized theory \cite{Shk17arx} of the hot-electron FFI has recently allowed us to fit a non-monotonic magnetic-field dependence of the instability velocity in Nb thin films with different pinning strengths \cite{Dob17arx}. A systematic experimental study of pinning effects on the LO instability in Nb films with nanogrooves is currently under way and will be reported in a forthcoming publication.

\section{Conclusion}
\label{sConclusion}
To sum up, the effect of pinning on self-heating triggering the LO flux-flow instability in superconducting thin films has been investigated theoretically. The problem was considered on the basis of the Bezuglyj-Shklovskij generalization of the LO theory, with an account for a finite heat removal from the quasiparticles at temperature $T^\ast$ to the bath at temperature $T_0$. The instability critical parameters, namely the current density $j^{\ast}$, the electric field $E^{\ast}$, the power density $P^{\ast}$, and the vortex velocity $v^{\ast}$ have been calculated and graphically analyzed as functions of the magnetic field value and the pinning strength. With increasing pinning strength at a fixed magnetic field value $E^{\ast}$ has been found to decrease, $j^{\ast}$ to increase, while $P^{\ast}$ and $T^{\ast}$ remain practically constant. An account for vortex pinning has substantially supplemented the well-established FFI physical picture. Vortex pinning may be the cause for eventual discrepancies between experiments on superconductors with strong pinning and the LO results including their subsequent refinements. The theoretical predictions for a decrease of the instability critical velocity and an increase of the instability current qualitatively agree with the results of electrical resistance measurements and computer simulations \cite{Sil12njp}. For a quantitative comparison of theory with experiment a series of samples with different nanostructure periods and pinning strengths is required, to systematically investigate the flux-flow instability in a broad range of magnetic fields and pinning intensities.

\section*{Acknowledgements}
OD acknowledges financial support through DFG Grant DO1511/3-1. Funding from the European Unions Horizon 2020 research and innovation program under Marie Sklodowska-Curie Grant Agreement No. 644348 (MagIC) is acknowledged.


%

\end{document}